\documentclass[12pt]{article}
\input epsf
\begin{document}

\title{Static solutions in the $U(1)$ gauged Skyrme model}
\author{B.M.A.G. Piette\thanks{e-mail address: B.M.A.G. Piette@durham.ac.uk}
\\
{\small Department of Mathematical Sciences}\\{\small University of
Durham}\\
{\small Durham DH1 3LE, UK}\\ \\{\small and}\\
D. H. Tchrakian\thanks{e-mail
address: tigran@thphys.may.ie}\\{\small Department of Mathematical
Physics}\\{\small National University of Ireland, Maynooth}\\
{\small Maynooth,
Ireland}\\{\small and}\\{\small School of Theoretical Physics}\\
{\small Dublin
Institute for Advanced Studies}\\{\small 10 Burlington Road}\\{\small
Dublin 4, Ireland}} \date{}

\newcommand{\dd}{\mbox{d}}\newcommand{\tr}{\mbox{tr}}
\newcommand{\ee}{\end{equation}}
\newcommand{\be}{\begin{equation}}
\newcommand{\ii}{\mbox{i}}\newcommand{\e}{\mbox{e}}
\newcommand{\pa}{\partial}\newcommand{\Om}{\Omega}
\newcommand{\vep}{\varepsilon}
\newcommand{\bfph}{{\bf \phi}}
\newcommand{\lm}{\lambda}
\def\theequation{\arabic{equation}}
\renewcommand{\thefootnote}{\fnsymbol{footnote}}
\renewcommand{\r}[1]{(\ref{#1})}
\newcommand{\bfR}{{\sf R\hspace*{-0.9ex}\rule{0.15ex}%
{1.5ex}\hspace*{0.9ex}}}
\newcommand{\N}{{\sf N\hspace*{-1.0ex}\rule{0.15ex}%
{1.3ex}\hspace*{1.0ex}}}
\newcommand{\Q}{{\sf Q\hspace*{-1.1ex}\rule{0.15ex}%
{1.5ex}\hspace*{1.1ex}}}
\newcommand{\C}{{\sf C\hspace*{-0.9ex}\rule{0.15ex}%
{1.3ex}\hspace*{0.9ex}}}
\renewcommand{\thefootnote}{\arabic{footnote}}

\newcommand{\ie}{{\it i.e.}}

\newcommand{\ble}{\begin{array}[b]{l l}}
\newcommand{\ele}{\end{array}}
\newcommand{\blle}[1]{\begin{array}[b]{l l l}}
\newcommand{\elle}{\end{array}}

\maketitle

\begin{abstract}
We use a prescription to gauge the $SU(2)$ Skyrme model with a 
$U(1)$ field, characterised by a conserved
Baryonic current. This model reverts to the usual Skyrme model in the
limit of the gauge coupling constant vanishing. We show that there exist
axially symmetric static solutions with zero magnetic charge, which can be
electrically either charged or uncharged. 
The energies of the (uncharged) gauged Skyrmions are less than the energy
of the (usual) ungauged Skyrmion. For physical values of the parameters
the impact of the $U(1)$ field is very small, so that it can be treated as
a perturbation to the (ungauged) spherically symmetric Hedgehog. This
allows the perturbative calculation of the magnetic moment.
\end{abstract} \medskip
\medskip

\section{Introduction}
For a long time now, much attention has been paid to the Skyrme \cite{S} 
model in 3 dimensions. It is believed to be an effective theory for
nucleons in the large $N$ limit of QCD at low energies. The classical
properties as well as the quantum properties of the model are in
relatively good agreement with the observed properties of small nuclei
\cite{W,ANW,BBT}.

Gauged Skyrme models have been used in the past. The $U(1)$ gauged
model~\cite{W,CW} was used to study the decay of nucleons in the vicinity
of a monopole \cite{CW}, while the $SU(2)_L$ gauged model~\cite{DF}
was used to study the decay of nucleons when the Skyrme model is coupled to
the weak interactions~\cite{DF}. The
Skyrme model has also been used to compute the quantum
properties of the Skyrmion \cite{ANW} where the gauge degrees of freedom
are quantised to compute the low energy eigenstates of a Skyrmion.
These states were identified as the proton, the neutron and the delta.

The aim of this work is to show that the Skyrme model can be coupled
to a self contained electromagnetic field and that this $U(1)$ gauged
model has stable classical solutions. In addition to these solitons
with vanishing magnetic and electric flux, we show that
this system supports solutions with nonvanishing electric
flux which are analogous to the dyon solutions of the Georgi-Glashow
model, just as the uncharged solitons are the analogues of the
monoploles~\cite{tp} of that model. The electrically charged lumps have
larger energy, or mass, than the uncharged soliton, just like the Julia-Zee
dyon~\cite{JZ} has larger energy, or is heavier, than the (electrically
uncharged) monopole. We shall
refer to these lumps as {\it charged $U(1)$ Skyrmions}.

In addition to its intrinsic
interest as a soliton in the Maxwell gauged Skyrme model, the present
work is also an example of a soliton in a $d$-dimensional $SO(N)$
gauged $S^d$ model with $N<d$ for the case $d=3,\: N=2$ , extending
the results of Ref.~\cite {T} which were restricted to the $N=d$ cases.
(The work of Ref.~\cite{T} consists of establishing topological lower
bounds for the generic case, encompassing earlier examples in
two~\cite{SGPS} and three~\cite{F,AT} dimensions respectively.)
The gauging prescription used here by us coincides precisely with
that used in Ref.~\cite{CW} and permits the establishing of a
topological lower bound which did not feature in Ref.~\cite{CW} and
which is carried out here to establish the stability of the soliton.
Such lower bounds are absent in the other prescription 
of gauging the Skyrme model as in Refs.~\cite{DF}. (Notice that we
name the sigma models after the manifold in which the fields take their
values rather than using the name of the
symmetry group for the model. Thus
what is sometime called the $O(d+1)$ model in the literature will be
refered to as the $S^d$ model.)

The $U(1)$ gauged $SU(2)$ Skyrme model is described by the Lagrangian 
\cite{CW}
$$
\label{LagU}
{\cal L}
= {F_{\pi}^2 \over 16} Tr\Bigl(D_{\mu}U D_{\mu}U^{\dag} \Bigr)
- { 1 \over 32 a^2}
  Tr \Bigl ([ (D_{\mu}U) U^{\dag}, (D_{\mu}U) U^{\dag}] \Bigr )^2
- {1\over 4} {\cal F}_{\mu \nu}^2 
\label{Lag}
$$
where the $U(1)$ gauge covariant derivative is
\be
\label{1}
D_{\mu}U=\pa_{\mu} U +i e {\cal A}_{\mu}[Q,U] \: ,
\ee
where 
${\cal F}_{\mu \nu} = \partial_{\mu}{\cal A}_\nu - \partial_{\nu}{\cal A}_\mu$
and where the charge matrix of the quarks is expressed as
$Q={1\over 2}({1\over 3}1+\sigma_3)$. This differs from the covariant
derivative of Ref.~\cite{CW} only in the unimportant matter of the sign
of $i$ in \r{1}, which we have chosen for consistency of the convention
used in Ref.~\cite{T}. 

In what follows it will be more convenient  \cite{T} to parametrise the 
Skyrme field as an $S^3$ valued field
$\phi^a =(\phi^{\alpha},\phi^A)$,
$\alpha =1,2$, $A=3,4$ subject to the constraint $|\phi^a|^2 =1$. 
The two fields $U$ and $\phi$ are related to each other via the following
expression
\be
\label{2}
U=\phi^a \tau^a ,\qquad \qquad
U^{-1}=U^{\dagger}=\phi^a \tilde \tau^a
\ee
where $\tau^a =(i\sigma^{\alpha},i\sigma^3 ,1)$ and $\tilde \tau^a
=(-i \sigma^{\alpha},-i\sigma^3 ,1)$, in terms of the Pauli matrices
$(\sigma^1 ,\sigma^2 ,\sigma^3)$.

The gauge covariant derivative now can be re-expressed as
\be
\label{3}
D_{\mu} \phi^{\alpha} =\pa_{\mu} \phi^{\alpha}
+A_{\mu} \vep^{\alpha \beta}
\phi^{\beta} ,
\qquad D_{\mu} \phi^A =\pa_{\mu} \phi^A.
\ee
where $A_\mu = e {\cal A}_\mu$ and $F_{\mu \nu} = e {\cal F}_{\mu \nu}$.

The Lagrangian for the $U(1)$ gauged Skyrme model can then be written as
\be
\label{4}
{\cal L}=-\lambda_0 F_{\mu \nu}^2 + \lambda_1 |D_{\mu} \phi^a|^2
-\lambda_2 |D_{[\mu} \phi^a D_{\nu ]}\phi^b|^2 
\ee 
where the square brackets on the indices imply (total) antisymmetrisation
and where $\lambda_0^{-1} = 4 e^2$, $\lambda_1 = F_{\pi}^2 /8$ and 
$\lambda_2^{-1} = 8 a^2$. The late Greek indices $\mu$
label the Minkowskian coordinates, while the early Greek indices $\alpha
=1,2$ and the upper case Latin indices $A=3,4$ label the fields $\phi^a =
(\phi^{\alpha},\phi^A)$.

The static Hamiltonian pertaining to the Lagrangian \r{4} is
\be
\label{5}
\begin{array}{rl}
{\cal H}&=\lambda_0 F_{ij}^2 + \lambda_1 |D_i \phi^a|^2 +
\lambda_2 |D_{[i} \phi^a D_{j ]}\phi^b|^2 \\ \\
&+i2\lambda_0 |\pa_i A_0|^2 +A_0^2 \{ \lambda_1
|\phi^{\alpha}|^2
+16\lambda_2 [|\phi^{\alpha}|^2 |\pa_i \phi^A|^2 +{1\over 4}
|\pa_i(|\phi^A|^2)|^2] \},
\end{array}
\ee
where the indices $i=\alpha,3$ label the space-like coordinates.

To find the static solutions of the model, one would usualy solve the Euler 
Lagrange equations which minimise the Hamiltonian \r{5}, but because of the 
electric potential $A_0$, one must solve the Euler Lagrange equations derived 
from the Lagrangian \r{4}. We then look for static solutions, but, as for
the Julia-Zee dyon~\cite{JZ}, we have to impose the proper asymptotic 
behaviour for the electric potential to obtain static solutions which are 
electrically charged (in the classical sense, {\it i.e.} solutions where the 
flux of the electric field is non zero).

When the full equations of motion are written down, one finds as expected that 
there are static solutions for which $A_0 = 0$, {\it i.e.} solutions for which
the electric field is identically zero. For these solutions in the
temporal gauge, the equations of motion
reduce to the equations obtained by minimising the Hamiltonian \r{5}.
We study the solutions of {\it unit Baryon charge} 
of the $U(1)$ gauged Skyrme model with and whithout an electric field,
for various values of the $U(1)$ coupling constant (or equivalently the
Skyrme coupling). For physical values of these parameters in the model, we
find that the energy (mass) of the gauged Skyrmion does not differ
significantly from that of the ungauged charged-$1$ Skyrmion, namely the
familiar hedgehog~\cite{S}. This implies that for these values of the physical
parameters, the $U(1)$ gauged Skyrmion can be regarded as a perturbation
of the (ungauged) hedgehog, enabling the computation of the magnetic moments
of the gauged Skyrmion (i.e. the Neutron) and the shift of the energy of the
gauged Skyrmion away from the energy of the hedgehog~\cite{S}, perturbatively
using the method employed by Klinkhamer and Manton~\cite{KM} for the
sphaleron of the Weinberg--Salam model.

In Section {\bf 2}, we define the topological charge and establish the
corresponding lower bound on the energy functional. In Section {\bf 3}
we present the solutions which have no electric fields in the
first subsection and electrically charged solutions in the second
subsection. The perturbation analysis of the gauged Skyrmion around the
(ungauged) hedgehog is carried out in Section {\bf 4}, and
Section {\bf 5} is devoted to a discussion of our results.

\section{The topological charge and lower bound}

The definition of the topological charge is based on the criterion that
it be equal to the Baryon number, namely the degree of the map. For the
gauged theory however, this quantity must be {\it gauge invariant} as well. 
This requirement can be systematically~\cite{T} satisfied by
arranging the gauge invariant topological charge density to be the sum of
the usual, {\it gauge variant} winding number density
\be
\label{6}
\varrho_0 =\vep_{ijk} \vep^{abcd} \pa_i \phi^a \pa_j \phi^b \pa_k
\phi^c \phi^d \: ,
\ee
plus a total divergence whose surface integral vanishes due to the finite
energy conditions, such that the combined density is gauge invariant. In
3 dimensions, this is given explicitly in Refs.~\cite{T,AT} for the
$SO(3)$ gauged $S^3$ model, and for the present case of interest, namely
the $SO(2)$ gauged $S^3$ model, the charge density can be derived from
that of the $SO(3)$ gauged model by contraction of the gauge group $SO(3)$
down to $SO(2)$. It can also be arrived at directly. To state the
definition of the charge, we denote the gauge covariant counterpart of
\r{6} by
\be
\label{7}
\varrho_G =\vep_{ijk} \vep^{abcd} D_i \phi^a D_j \phi^b D_k
\phi^c \phi^d \: ,
\ee
so that using the notations \r{6} and \r{7} we have the definition of the
gauge invariant topological charge
\begin{eqnarray}
\label{8a}
\varrho
&=& \: \varrho_0 +\pa_i \Omega_i , \\
\label{8b}
&=& \varrho_G +\frac{3}{2} \vep_{ijk} F_{ij} (\vep^{AB} \phi^B D_k
\phi^A)\: .
\end{eqnarray}
In \r{8a} the density $\Omega_i$ is the following gauge {\it variant} form
\be
\label{9}
\Omega_i =3\vep_{ijk} \vep^{AB} A_j \pa_k \phi^A \: \phi^B.
\ee
The flux of $\Omega_i$ vanishes, as can deduced by anticipating the finite
energy conditions to be stated later.

Note that the 3-volume integral of $\varrho_0$ in \r{8a} is the degree of the 
map for the ungauged system namely the Baryon number.

Identifying $\varrho$, \r{8b}, with the naught component $j^0$ of the Baryon
current, $j^{\mu}$ is defined by
\begin{eqnarray}
\label{10a}
j^{\mu}
&=&
\vep^{\mu \nu \rho \sigma} \vep_{abcd} \pa_{\nu}\phi^a
\pa_{\rho}\phi^b \pa_{\sigma}\phi^c \phi^d
+3 \: \vep^{\mu \nu \rho \sigma} \: \pa_{\nu} (A_{\rho} \vep^{AB}
\phi^B \pa_{\sigma} \phi^A) \\
&=&
\label{10b}
\vep^{\mu \nu \rho \sigma} \: \vep_{abcd} D_{\nu}\phi^a
\: D_{\rho}\phi^b D_{\sigma}\phi^c \phi^d
-\frac{3}{2} \: \vep^{\mu \nu \rho \sigma} \:  F_{\nu \rho}
(\vep_{AB}\phi^B D_{\sigma}
\phi^A).
\end{eqnarray}

The 4-divergence of \r{10a} receives a contribution only from its first term,
which being locally a total divergence implies that the 3-volume integral 
of $j^0$ is a conserved quantity.
Alternatively we consider the 4-divergence of \r{10b}, 
\be
\label{11}
\pa_{\mu}j^{\mu} =6\vep^{\mu \nu \rho \sigma} \vep_{\alpha \beta} \vep_{AB}
D_{\mu}\phi^{\alpha} D_{\nu}\phi^{\beta} D_{\rho}\phi^A D_{\sigma}\phi^B
\ee
which is analogous to the corresponding quantity in the work of Goldstone and 
Wilczek \cite{GW}. This contrasts with the expression for the total 
divergence of the topological current in the work of d'Hoker and Farhi 
\cite{DF}, where a different gauging prescription is used leading to that 
quantity being equal to the local anomaly.

We now proceed to find a model whose Hamiltonian ${\cal H}_0$ is
bounded from below by the topological charge density defined by \r{8b}. We
will then show that the Hamiltonian \r{5} is given by ${\cal H}_0$
plus certain positive definite terms.

First of all, we reproduce the density $\varrho_G$,
\r{7}, in \r{8b} by using the following inequality
\be
\label{12}
(\kappa_3 D_i\phi^a -\vep_{ijk} \vep^{abcd} \kappa_2^2 D_j \phi^b D_k
\phi^c \phi^d)^2 \ge 0
\ee
where the two constants $\kappa_3$ and $\kappa_2$ have the dimensions of
length. Expanding the square, we get $\varrho_G$ on the right hand side of
\be
\label{13}
\kappa_3^2 (D_i \phi^a)^2 +\kappa_2^4 (D_{[i} \phi^a D_{j ]}\phi^b)^2 \ge
2\kappa_3 \kappa_2^2 \varrho_G.
\ee

To reproduce the other term in \r{8b}, $\frac{3}{2} \vep_{ijk}
F_{ij}(\vep^{AB} \phi^B \pa_{k} \phi^A)$, we use the following
inequality
\be \label{14}
(\kappa_0^2 F_{ij} -{1\over 2}\kappa_4 \vep_{ijk} \vep^{AB} \phi^B D_k
\phi^A)^2 \ge 0
\ee
yielding
\be
\label{15}
\kappa_0^4 F_{ij}^2 +\kappa_4^2 {1\over 4}(\vep^{AB} \phi^B
D_{i} \phi^A)^2 \ge \kappa_0^2 \kappa_4 \vep_{ijk} F_{ij} (\vep^{AB}
\phi^B D_k \phi^A) .
\ee

With the special choice for the relative values of the constants
$3\kappa_3 \kappa_2^2 =\kappa_4 \kappa_0^2$, the sum of \r{11} and \r{13} 
yields the following
\be
\label{16}
\kappa_0^4 F_{ij}^2 +
\kappa_3^2 (D_i \phi^a)^2 +\kappa_2^4 (D_{[i} \phi^a D_{j ]}\phi^b)^2
+\frac{9 \kappa_3^2 \kappa_2^4}{4\kappa_0^4} (\vep^{AB} \phi^B D_{i}
\phi^A)^2 \ge 2\kappa_3 \kappa_2^2 \varrho .
\ee

The right hand side of \r{16} is now proportional to the topological
charge
density $\varrho$ defined by \r{8b} so that the inequality \r{16} can
be interpreted as the topological inequality giving the lower bound on the
energy density functional if we define the latter to be the left hand side
of \r{16}, namely
\be
\label{17}
{\cal H}_0 =\kappa_0^4 F_{ij}^2 +
\kappa_1^2 (D_i \phi^a)^2 +\kappa_2^4 (D_{[i} \phi^a D_{j ]}\phi^b)^2
+\frac{9 \kappa_3^2 \kappa_2^4}{4\kappa_0^4} (\vep^{AB} \phi^B D_{i}
\phi^A)^2.
\ee

The Hamiltonian system \r{17} is almost the Hamiltonian of the gauged
Skyrme model \r{5} (remember that $A_0 = 0$). It differs from the latter
only in its last term. Now we can use the identity
\be
\label{18}
(\vep^{AB} \phi^B D_i \phi^A)^2
=(D_i\phi^a)^2 -\bigg[{1\over 2}(\phi^{[\alpha} D_i\phi^{\beta ]})^2
+(\phi^{[\alpha} D_i \phi^{A]})^2 \bigg]
\ee
and add the positive definite term
$\frac{\kappa_3^2 \kappa_2^4}{9\kappa_0^4} \bigg[{1\over 2}(\phi^{[\alpha}
D_i\phi^{\beta ]})^2 +(\phi^{[\alpha} D_i \phi^{A]})^2 \bigg]$ appearing
on the right hand side of \r{18} to ${\cal H}_0$ in \r{17} to end up with
the Hamiltonian for the $U(1)$ gauged Skyrme model:
\be
\label{19}
{\cal H}=\kappa_0^4 F_{ij}^2 + \kappa_1^2 (D_i \phi^a)^2 +\kappa_2^4
(D_{[i} \phi^a D_{j ]}\phi^b)^2 \ge  2\kappa_3 \kappa_2^2 \varrho
\ee
which is nothing but the static Hamiltonian \r{5} in the temporal
gauge $A_0 =0$, and where
\be
\label{20}
\lambda_1 = \kappa_3^2 (1+\frac{9 \kappa_2^4}{4\kappa_0^4}),\qquad 
\lambda_0 = \kappa_0^4,\qquad 
\lambda_2 = \kappa_2^4.
\ee
By virtue of \r{16}, \r{19} is also bounded from below by $2\kappa_3
\kappa_2^2 \varrho$, namely by a number proportional to the topological
charge density $\varrho$.

We thus see that ${\cal H}_0$ can be considered as a minimal ($U(1)$
gauged) model, but from now on, we will restrict our attention to the
physically more relevant model \r{19} and integrate it numerically to find
its topologically stable finite energy solitons.

The soliton solutions to the system \r{19} can only be
found by solving the
second-order Euler-Lagrange equations, and {\it not} some first-order
Bogomol'nyi equations since saturating the inequalities \r{12} and \r{14}
would not saturate the lower bound on the energy density functional ${\cal
H}$. In this context we note that saturating \r{12} and \r{14} does indeed
saturate the topological lower bound on the functional ${\cal H}_0$ by
virtue of the inequality \r{16}, and should it have turned out that the
Bogomol'nyi equations arising from the saturation of \r{12} and \r{14}
supported non-trivial solutions, then ${\cal H}_0$ would have been a very
interesting system to consider. As it turns out however,
these Bogomol'nyi equations have only trivial solutions in exactly the same
way as in the case of the (ungauged) Skyrme model \cite{S}.

The energy for the static configuration, when the electric field vanishes, 
is expressed as
\be
\label{21}
E(\lambda_0,\lambda_1,\lambda_2) = \int d^3 x \:  [\lambda_0 F_{ij}^2 +
\lambda_1 (D_{i} \phi^a)^2
+\lambda_2 (D_{[i} \phi^a D_{j]}\phi^b)^2 ]
\ee
and performing the dilation $x
\rightarrow \sigma x$, $A_\mu \rightarrow \sigma^{-1} A_\mu$, we get
\be
\label{22}
E(\lambda_0,\lambda_1,\lambda_2) = \int d^3 x\: [{\lambda_0\over \sigma}
F_{ij}^2 + \sigma \lambda_1 (D_{i} \phi^a)^2 +{\lambda_2\over \sigma}
(D_{[i} \phi^a D_{j]}\phi^b)^2 ].
\ee
If we choose $\sigma = \bigl({\lambda_2 \over \lambda_1}\bigr)^{1/2}$ then
we have
\be
\label{23}
E(\lambda_0,\lambda_1,\lambda_2) =
(\lambda_1\lambda_2)^{1/2} E({\lambda_0\over\lambda_2},1,1)\: ,
\ee
from which we see that we can set $\lambda_1 = \lambda_2 = 1$
without any loss of
generality. By virtue of \r{19} and \r{23}, we can finally state
\be
\label{24}
E(\lambda_0,\lambda_1,\lambda_2) \ge
2 ({\lambda_1 \lambda_2\over 1 + {9 \lambda_2 \over 4\lambda_0}})^{1/2} \int
d^3 x\: \varrho.
\ee

Notice that for the usual Skyrme model we have
\be
\label{25}
\begin{array}{rl}
E_{sk}(\lambda_1,\lambda_2)
=& \int dx^3 [\lambda_1 (\partial_{i} \phi^a)^2 +\lambda_2 (\partial_{[i}
\phi^a \partial_{j]}\phi^b)^2 ]\\ \\ =& (\lambda_1\lambda_2)^{1/2}
E_{sk}(1,1)\\ \\ \ge& 2 ({\lambda_1 \lambda_2})^{1/2} \int dx^3
\varrho_0. \end{array}
\ee
We will use \r{25} to compare the numerical solutions of the gauged
Skyrme model with the solutions of the (ungauged) Skyrme model.

We would like to point out that the topological stability considerations 
discussed in this Section apply only to the solutions with no electric 
field, \ie with $A_0 = 0$.

\section{The soliton and the charged $U(1)$ Skyrmion}

To find the static solutions, we have to look for the largest
symmetry group of the functional to be subjected to the variational
principle, and look for solutions which are invariant under that symmetry
group. For the solutions in the $A_0 =0$ gauge this is the static 
Hamiltonian \r{19}, while for the solutions in the $A_0 \neq 0$ it is
the Lagrangian \r{4}. For our choice of gauge group the largest
symmetry is the $SO(2)$ group corresponding to an axial rotation in 
space-time and a gauge transformation on the gauge field. Defining the axial
variables $r=\sqrt {x_1^2 +x_2^2}$ and and $z=x_3$ in terms of the
coordinates $x_{i} =(x_{\alpha} ,x_3)$, $\alpha =1,2$, the most general
axially symmetric Ansatz \cite{BK}
for the fields $\phi^a =(\phi^{\alpha} ,\phi^A)$
(with $\alpha =1,2$ and $A =3, 4)$, and, $A_i =(A_{\alpha} ,A_3)$, is

\be
\label{1.1}
\phi^{\alpha} =\sin f \sin g \: \: n^{\alpha}\: ,\qquad \phi^3 =\sin f
\cos g \: , \qquad \phi^4 =\cos f \: ,
\ee
\be
\label{1.2}
A_{\alpha} =\frac{a(r,z) -n}{r}\ \vep_{\alpha \beta}\ \hat x_{\beta} + 
\frac{c_2(r,z)}{r} \hat x_{\alpha} 
\qquad A_3 = \frac{c_1(r,z)}{r},  \qquad A_0 = \frac{b(r,z)}{r},  
\ee
with $n^{\alpha} =(\sin n\phi ,\cos n\phi)$ in terms of the azimuthal
angle $\phi$ and $\hat x_{\alpha} ={x_{\alpha}/ r}$. $n$ in $n^{\alpha}$
is the {\it vorticity}, which for the Nucleons of interest to us here,
equals {\em one}, $n=1$. The
functions $a, b, c_1, c_2, f$ and $g$ both depend on $r$ and $z$.

Our Ansatz (\ref{1.2}) for the $U(1)$ field consists of decomposing the latter
in the most general tensor basis possible. We will find out below, when
we compute the Euler--Lagrange equations, that the functions $c_1$ and $c_2$
vanish identically. Anticipating this, we suppress them henceforth. In its
final form this Ansatz agrees with that used in Ref.~\cite{BK}, the latter
being arrived at by specialising the Rebbi--Rossi Ansatz for the axially
symmetric $SO(3)$ field.

The static Hamiltonian, \ie the $T_{00}$ component of the energy mometum tensor
$T_{\mu\nu}$, is then given by
\be\label{ham}
\begin{array}{rl}
H=&\int\bigg\{ {\lambda_0\over r^2}\bigg[a_r^2+ a_z^2
               +(b_r-{b\over r})^2 + b_z^2 \bigg] \\
\\
&+{\lambda_1\over 2} \bigg[ f_r^2 +f_z^2 +\sin^2 f (g_r^2 +g_z^2) 
    +{a^2 + b^2 \over r^2} \sin^2 f \sin^2 g \bigg] \\
\\
&+2 \lambda_2\sin^2 f \bigg[ (f_r g_z -f_z g_r)^2 +\sin^2 g 
[{a^2+b^2\over r^2} (f_r^2 +f_z^2+(g_r^2 +g_z^2) \sin^2 f )] 
\bigg] \bigg\} r dr dz.
\end{array}
\ee

The boundary conditions for the Skyrmion fields are the same as the boundary 
conditions for the hedgehog ansatz when expressed in the cylindrical 
coordinates where $g = \pi/2+\arctan(z/r)$ and defining $R = (z^2+r^2)^{1/2}$, 
$f \equiv f(R) $ with $f(0) = \pi$ and $\lim_{R\rightarrow\infty} f(R) = 0$.
From this we can deduce that the function $f$ has a fixed value at the 
origin and at infinity. For smothness along the $z$ axis, each field,
that is $f$, $g$ and $A_{\alpha}$ must  satisfy the condition that 
the partial derivative with respect to of the field at $r = 0$ vanishes.
The boundary conditions and the asymptotic behaviours for $a$ and $b$ are 
chosen so that the gauge fields $A_\mu$ are well defined and $A_0$ looks 
asymptoticaly like a coulomb field 
(i.e. with an electric charge but no magnetic charge). We also require that 
the total energy be finite. 
These conditions leads to the following constraints:

\be
\label{boundaries}
\begin{array}{rlrlrl}
f(0,0) &= \pi&  \qquad f(r \rightarrow \infty,z \rightarrow \infty) &= 0 & 
f_r(r=0,z) &= 0 \\
g(r=0,z < 0) &= 0& \qquad g(r=0,z > 0) &= \pi & 
   \qquad g_R \vert_{R\rightarrow\infty} &= 0\\
a(r=0,z) &= 1&  \qquad a_r \vert_\infty \ &= 0&
   \qquad a_z \vert_\infty\ &= 0 \\
a_r(r=0,z) & = 0 & A_0(r \rightarrow \infty,z \rightarrow \infty) &= V_0+q/r& 
 A_0(r=0,z) &= 0   
\end{array}
\ee
where
$R = (z^2+r^2)^{1/2}$ and where we have used the notation 
$\frac{\pa a}{\pa r} =a_r$ etc. Note that the field $g$ is undefined 
at the origin and the resulting discontinuity of $g$ at that point is  an 
artefact of the coordinate system used. The asymptotic behaviour of
$A_0$  at infinity will be discussed in a later Section. To solve the 
equations numerically, it is more conveniant to use the field $A_0$ rather 
than $b$; this is why we have expressed the boundary condition in terms of 
that field. On the other hand, the equations take a simpler form when written
in term of $b$, so we shall still use it below.

Now the volume integral (with the appropriate normalisation of $12\pi^2$)
of $\varrho_0$ given by \r{6} is the degree of the map, or, the Baryon
number. It is straightforward to verify that when the Ansatz \r{1.1} is
substituted in $\varrho_0$ and the volume integral is computed subject
to the boundary conditions given above, the result will equal the
integer $n$ defined in \r{1.1}. Thus, the Baryon number of the field
configuration \r{1.1} equals the vortex number $n$.
In what follows, we will restrict ourselves to {\it unit} Baryon number,
$n=1$, \ie   to the Nucleons. 

Before we proceed to substitute the Anstaz \r{1.1}, \r{1.2} into the field
equations, we calculate the Baryon current \r{10a} for the field configurations
\r{1.1}, \r{1.2} described by the solutions we seek. We express the space-like
part of this current $j_i$ in the radial direction flowing out of the normal
to the surface of the cylinder which we denote by $j_r$, and in the $z$ 
direction which we denote by $j_z$. The result is
\begin{eqnarray}
\label{radial}
j_r &=& {6\over r}
[(\dot{f} g_z -\dot{g} f_z)a \sin^2 f \sin g \\ \nonumber
&&
+{1\over 2}(\dot{g} a_z -\dot{a} g_z)\sin f \cos f \sin g
+{1\over 2}(\dot{a} f_z -\dot{f} a_z)\cos g]  \\\label{z}
j_z &=& -{6\over r}
[(\dot{f} g_r -\dot{g} f_r)a \sin^2 f \sin g \\ \nonumber
&&
+{1\over 2}(\dot{g} a_r -\dot{a} g_r)\sin f \cos f \sin g
+{1\over 2}(\dot{a} f_r -\dot{f} a_r)\cos g] \: ,
\end{eqnarray}
where we have denoted $\frac{\pa f}{\pa t} =\dot{f}$ etc. Note that the
Baryon current \r{radial}, \r{z} are {\it not} sensitive to the charge
of the Nucleon, {\it i.e.} that the function $b(r,z)$ does not feature in them.

We now turn to the equations to be solved, namely the Euler-Lagrange
equations arising from
the variational principle applied to the Lagrangian \r{4}, in the
{\it static limit}. Substituting the ansatz \r{1.1} \r{1.2} into these 
equations of motion leads to 
 \be\ble
a_{rr}& - {a_r\over r} - a_{zz}
  - a \sin^2(f) \sin^2(g)
\Bigl[{\lambda_1 \over 2\lambda_0}
  + { 2 \lambda_2\over\lambda_0}[f_r^2+f_z^2 + (g_r^2+g_z^2) \sin^2(f)]
\Bigr] = 0\\
&\\
b_{rr}&- {b_{r}\over r}+{b\over r^2}+b_{zz} 
  - b \sin^2(f)\sin^2(g) \Bigl[ {\lambda_1 \over 2\lambda_0} 
  + {2 \lambda_2\over\lambda_0}[f_r^2+f_z^2 + (g_r^2+g_z^2) \sin^2(f)]
\Bigr] = 0\\
&\\
\Delta f& - (g_r^2 +g_z^2 + \frac{a^2 -b^2}{r^2} \sin^2g)\sin f \cos f +\\
+4 {\lambda_2 \over \lambda_1}& \sin f\Bigl[
  \bigr(g_r (f_{zz} g_r - f_r g_{zz} + f_z g_{rz} - f_{rz} g_z - f_z/r g_z)\\
  &     +g_z (f_{rr} g_z - f_z g_{rr} + f_r g_{rz} - f_{rz} g_r + f_r/r g_z)
  \bigl)\sin f  +(f_r g_z-g_r f_z)^2 \cos f\\
  &+\sin g /r^2 \Bigl( (a^2-b^2)
      \bigl(f_r^2+f_z^2- 2 (g_r^2+g_z^2) \sin^2 f\bigr) \cos f \sin g\\
  &\quad+\bigl((a^2-b^2) (f_{rr}+f_{zz}-f_r/r)
              +2 f_r (a a_r-b b_r) +2 f_z (a a_z-b b_z)
       \bigr)\sin f \sin g\\ 
    &\quad +2 (a^2-b^2)\bigl(f_r g_r+f_z g_z\bigr)\sin f \cos g 
     \Bigr)
\Bigr]= 0.\\
&\\
\Delta g& +2(f_r g_r +f_z g_z)\cot f 
                         -\frac{a^2 -b^2}{r^2} \sin g \cos g+ \\
+4 {\lambda_2\over\lambda_1} &\Bigl[
   f_z (f_z g_{rr}- f_{rr} g_z + f_{rz} g_r - f_r g_{rz} + f_z g_r/r)\\
 &+f_r (f_r g_{zz}- f_{zz} g_r + f_{rz} g_z - f_z g_{rz} - f_z g_z/r)\\
 &+\sin g/r^2 \Bigl((a^2-b^2) ((g_r^2+g_z^2)\sin^2 f-f_r^2-f_z^2)\cos g\\
         &\quad\bigl((a^2-b^2) (g_{rr}+g_{zz}-g_r/r)+2 g_r (a a_r - b b_r)
                          +2 g_z (a a_z - b b_z)
        \bigr) \sin^2 f \sin g\\
        &\quad+4(a^2-b^2) (f_r g_r+f_z g_z)\cos f \sin f \sin g
       \Bigr)
\Bigr]= 0.
\ele\label{eqgsk}
\ee

In the case of the $A_0 =0$ gauge, equations \r{eqgsk}
coincide with the Euler-Lagrange equations derived from the positive
definite Hamiltonian density \r{5}. Moreover in that case, those equations
also coincide with the Euler-Lagrange equations of the reduced two
dimensional Hamiltonian obtained by subjecting \r{5} to axial symmetry
by substituting the Ansatz \r{1.1}-\r{1.2} into it. This is expected due
to the strict imposition of symmetry. In the $A_0 \neq 0$ gauge, the
Euler-Lagrange equations are derived from the Lagrangian \r{4} which is
not positive definite. Nonetheless these equations coincide with those
arising from the reduced two dimensional Lagrangian obtained by subjecting
the Lagrangian \r{4} to axial symmetry. (This happens also for the Julia Zee
dyon \cite{JZ}.)


\subsection{$A_0 =0$ : $U(1)$ Skyrme soliton}

It is easy to see from \r{eqgsk} that there are solutions for which 
$b=0$ (\ie \ $A_0 =0$). As mentioned before, in that case, equation
\r{eqgsk} can be obtained by minimising the Hamiltonian \r{ham}.
Notice also that setting $a = 0$ is not compatible with our boundary 
conditions ($A_i$ would not be well defined at the origin). We thus expect our 
gauged solution to carry a non-zero magnetic field. 

To show this we have to solve equations \r{eqgsk} numerically for
the non-vanishing functions $f(r,z), g(r,z)$ and $a(r,z)$.
 
We have restricted our numerical integrations to the case where the vortex
number $n$ appearing in the axially symmetric Ansatz \r{1.1} is equal to
$1$, \ie our soliton carries unit Baryon number.

Using \r{ham}, we have found numerically that $E(1,1,1) = 24 \pi^2 1.01 $
whereas for those values of $\lambda_0, \lambda_1, \lambda_2$ the lower bound 
for the energy given by \r{24} is $ 24 \pi^2 0.555 $. In Figure 1, we present 
the total energy for the gauged
Skyrmion as a function of $\lambda_0$, together with the lower bound given
by \r{24}. Note that the asymptotic value of $E(\lambda_0 ,1,1)$ is 
$24 \pi^2 1.232$ as $\lambda_0 \to \infty$. As a comparison, the energy 
\r{25} for the ungauged Skyrmion is $E_{sk}(1,1) = 24 \pi^2 1.232$ with a 
lower bound set
at $24 \pi^2$. We see that $E(\lambda_0 =\infty ,1,1)=E_{sk}(1,1)$ which
means that as $\lambda_0 \rightarrow \infty$, the gauge coupling
$1/\lambda_0^{1/2}$ goes to zero and the gauged Skyrmion becomes in this
limit the ungauged Skyrmion.

It is interesting to note that the energy of gauged Skyrmion is smaller 
than the energy of the ungauged Skyrmion, as expected, but that on the other 
hand, the amount by which the energy of the gauged Skyrmion exceeds its 
topological lower bound is larger than the excess of the energy of the 
ungauged Skyrmion above its respective topological lower bound. For
example we can clearly see from Figure 1.a that at $\lambda_0=20$,
the energy of the gauged Skyrmion
$1.22$ (in units of $24\pi^2$) exceeds the lower bound $0.95$ by $0.27$.
This is larger than $0.232$, the excess of the ungauged Skyrmion
energy over its lower bound. For smaller values of $\lambda_0$ Figure 1.a
shows that the excess of the energy of the gauged Skyrmion over its lower
bound is even larger, hence this is a general feature.

In Figure 1.b, we also see that the Maxwell Energy, \ie the term 
proportional to $\lambda_0$ in (\ref{ham}), is decreasing as  $\lambda_0$
increases. Notice that we could have used for the Maxwell Energy the sum of all
the terms involving the gauge field functions $a$ and $b$ in (\ref{ham}), 
but this would lead to a figure similar to Figure 1.b.

\begin{figure*}[tp]
\unitlength1cm
\hfil\begin{picture}(6,8)
\epsfxsize=6cm
\epsffile{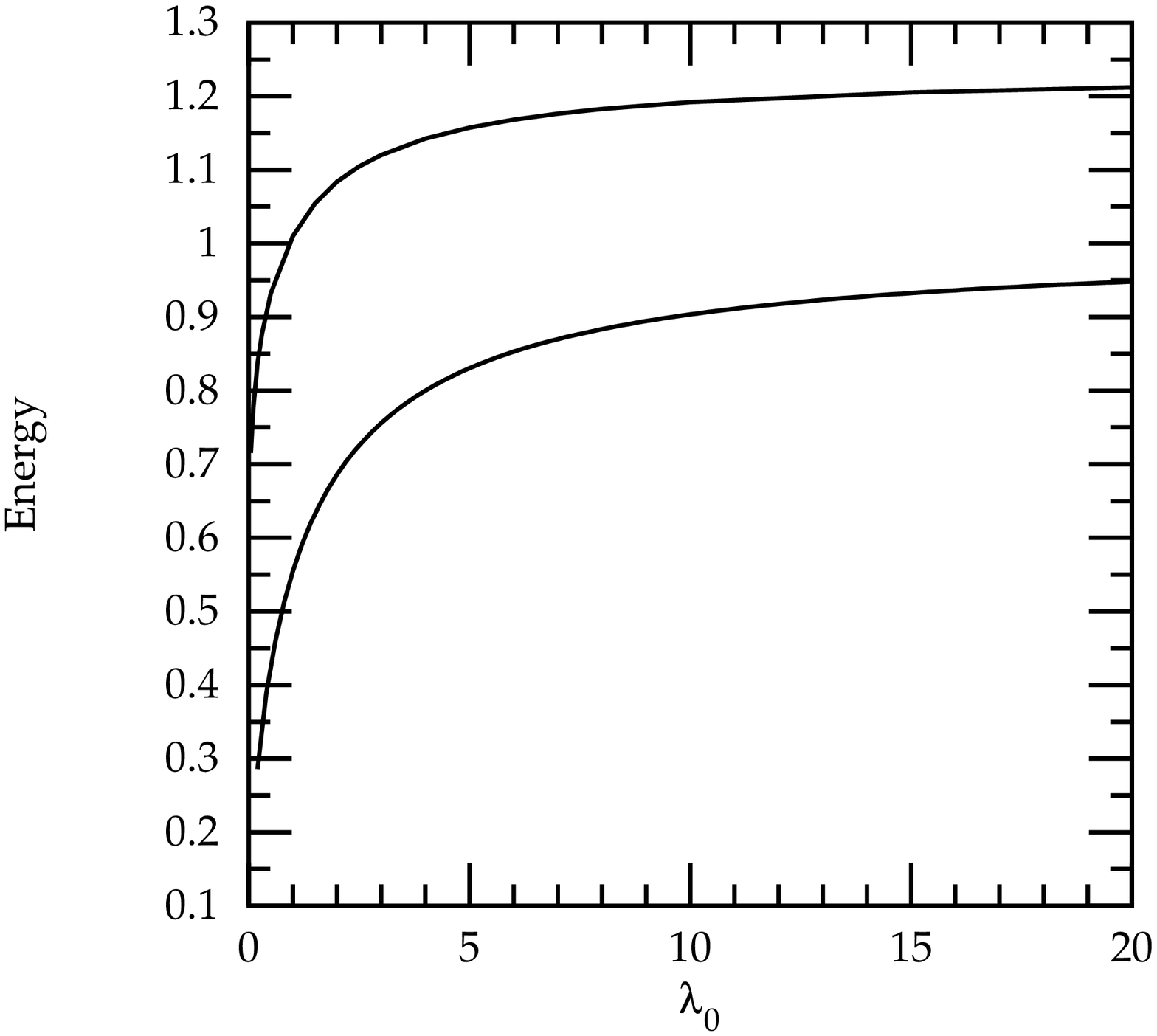}
\end{picture}
\hfil\begin{picture}(6,8)
\epsfxsize=6cm
\epsffile{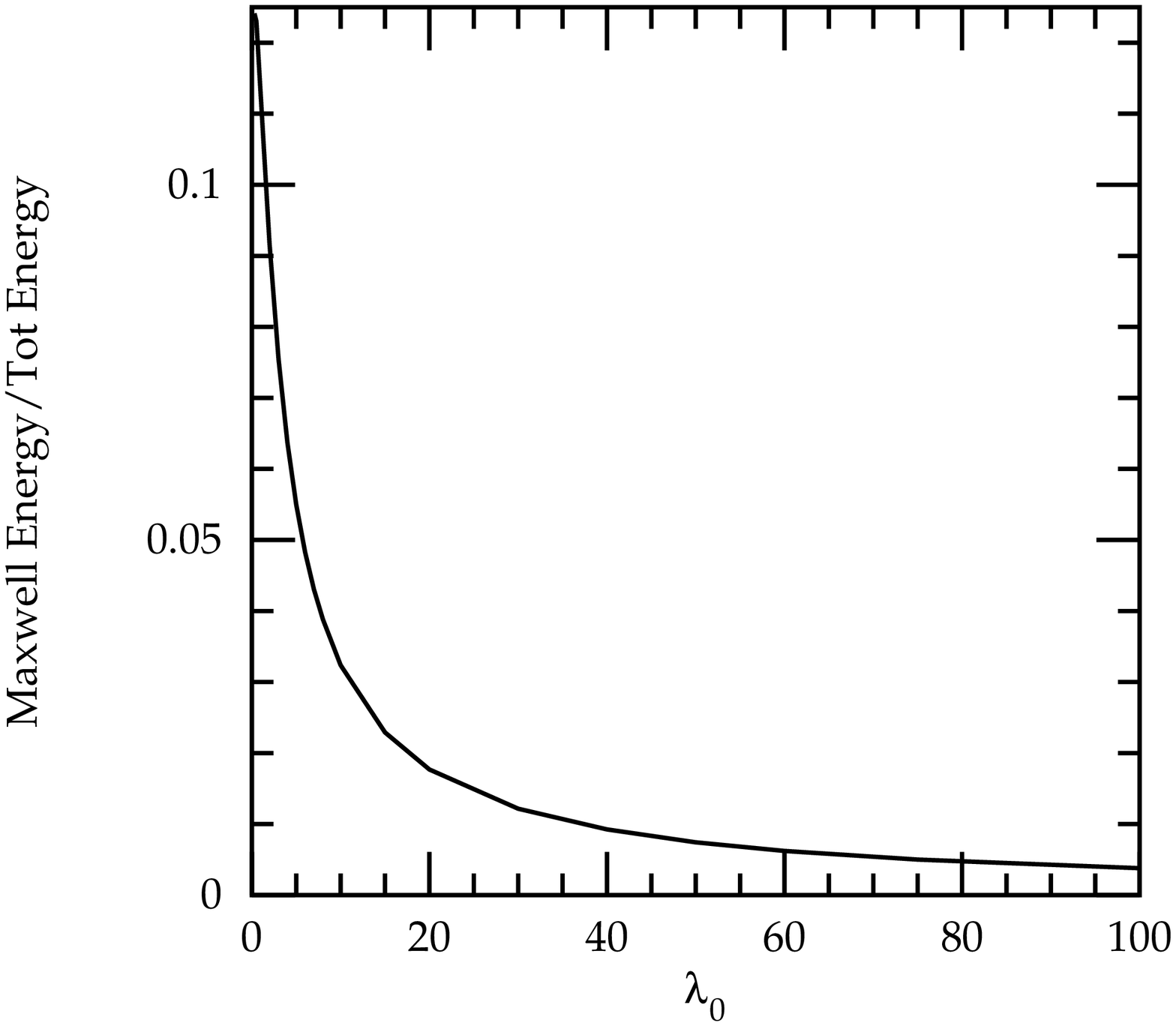}
\end{picture}
\caption{a) Energy and topological bound of the gauged Skyrmion in units of 
$24 \pi^2$. b) Ratio of the electromagnetic and the total energy as a function 
of $\lambda_0$ .}
\end{figure*}

In Figure 2, we show the profile and the level curve for
the energy density of the Skyrmion in the $r,z$ plane for $\lambda_0 = 1$. 
One sees clearly that the effect of the gauged field is to make the 
Skyrmion elongated along the $z$ axis. The magnetic field vectors of the
Skyrmion are parallel to the $r,z$ plane. In Figure 3, we show the
configuration of magnetic field using arrows to represent the magnetic
field vector at each point on the grid. Notice that there is a vortex
around the point $r=2, z = 0$. The magnetic field is thus generated by a
current flowing on a ring centred around the $z$ axis.

\begin{figure*}[pt]
\unitlength1cm
\hfil\begin{picture}(6,8)
\epsfxsize=6cm
\epsffile{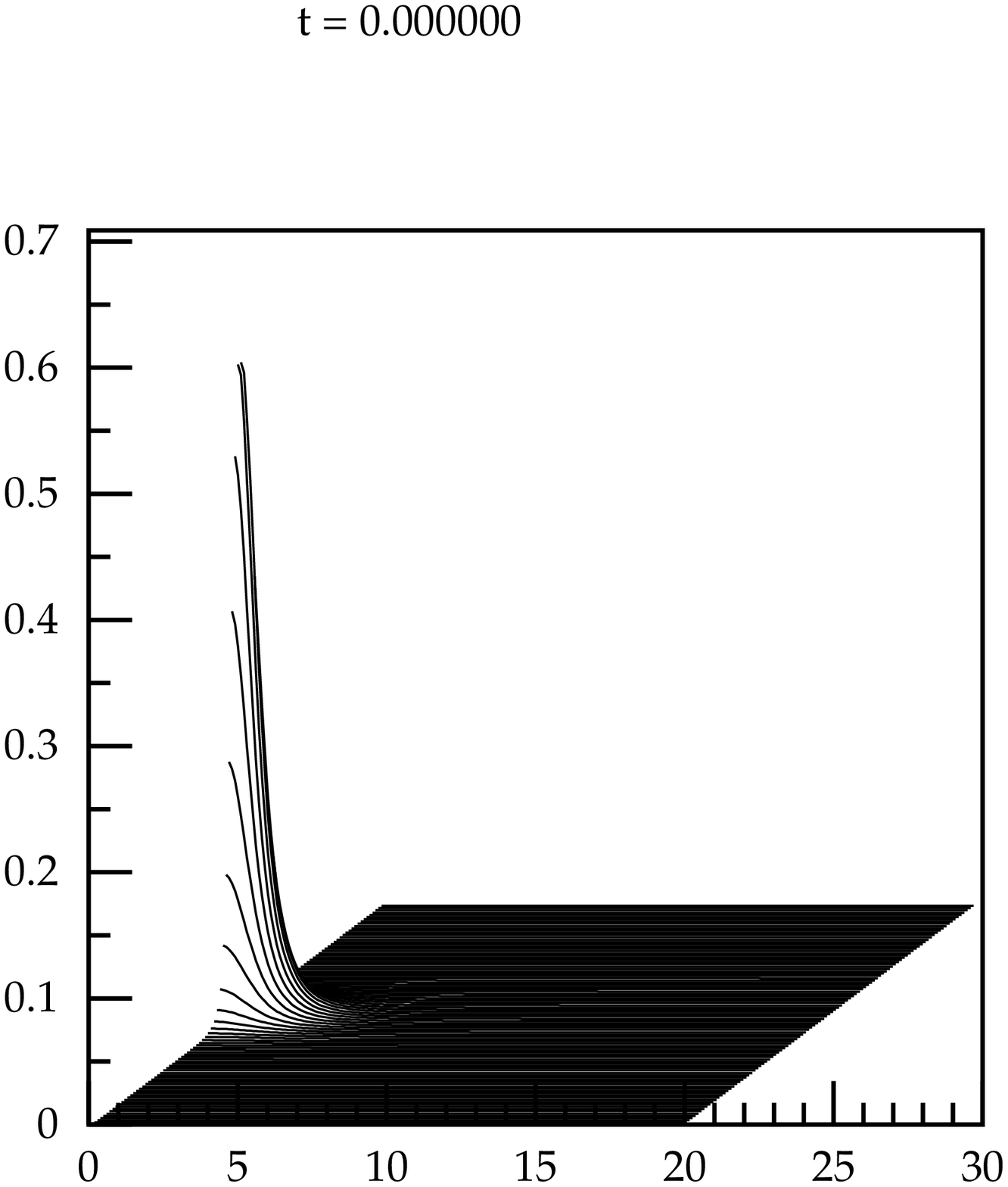}
\end{picture}
\hfil\begin{picture}(6,8)
\epsfxsize=6cm
\epsffile{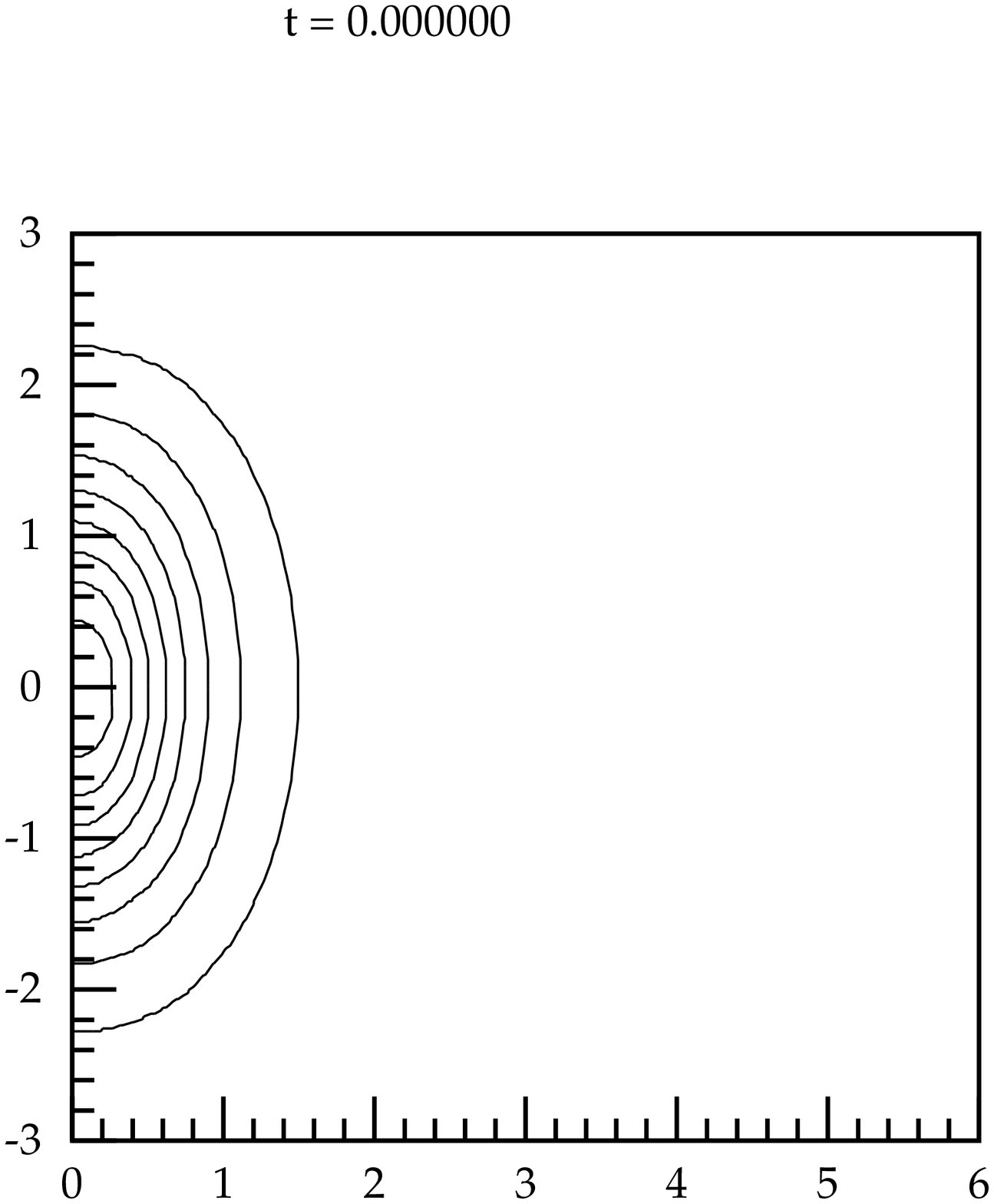}
\end{picture}
\caption{a) Energy density for the gauged Skyrmion in the
$(r,z)$ plane ($\lambda_0=\lambda_1=\lambda_2=1$). 
b) Energy density level curve for the gauged Skyrmion 
($\lambda_0=\lambda_1=\lambda_2=1$).}
\end{figure*}

\begin{figure}[ht]
\unitlength1cm
\hfil\begin{picture}(6,8)
\epsfxsize=6cm
\epsffile{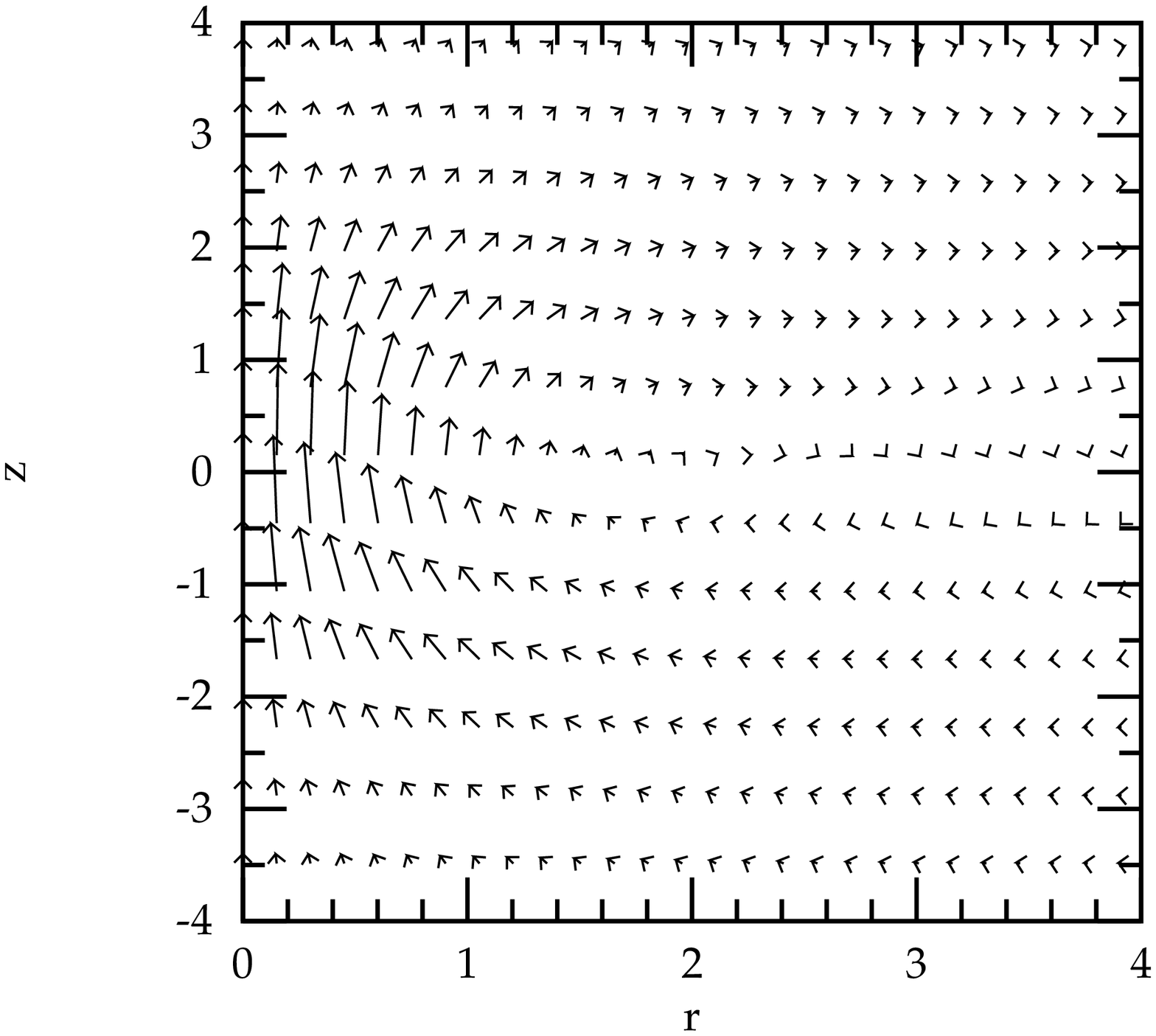}
\end{picture}
\caption{Magnetic Field of the gauged Skyrmion 
($\lambda_0=\lambda_1 =\lambda_2=1$).}
\end{figure}

In terms of the usual physical constants\cite{ANW}, we have
$\lambda_0^{-1} = 4 e^2$, $\lambda_1 = F_{\pi}^2 / 8$ and $\lambda_2^{-1}
= 8 a^2$ where we
use $a$ instead of the traditional $e$ for the Skyrme coefficient to avoid
confusion with the electric charge.

In our units, $c = \hbar = 1$, we have $e = (4 \pi \alpha)^{1/2}$ where
$\alpha = 1/137$ is the fine
structure constant. Choosing $F_{\pi} = 186 MeV$, we can find the value for
$a$ by requiring that the energy of the neutron $M_n = 939 MeV$
matches the energy of the Skyrmion:
$$
M_n = {F_{\pi} \over { 8 a}} E({2 a^2\over e^2},1,1).
$$
In Figure 1.a, we can
read the value of $E(\lambda_0,1,1)$ (given in units of $24 \pi^2 MeV$)
with $\lambda_0 = 2 {a^2/ e^2}$. We now have to find the value of $\lambda_0$ 
for which
$24 \pi^2 E(\lambda_0,1,1) = 2 (2\lambda_0)^{1/2} e M_n / F_{\pi}$. 
The intersection between the curve 
${e M_n \over 6 \pi^2 F_{\pi}} (2\lambda_0)^{1/2} $
and the curve $E(\lambda_0,1,1)$ in Figure 1.a
is located in the region where the
energy is virtually equal to the asymptotic value $E(\lambda_0,1,1) =
1.232$. This means that $a \approx 3 \pi^2 1.232 F_{\pi}/ M_n \approx
7.2$ and that $\lambda_0 \approx 1138$. We can thus conclude that the
effective impact of the Maxwell term we have added to the Skyrme model
is relatively small.

This justifies the procedure used in \cite{CW} where the Skyrmion was
coupled with an external magnetic field of a magnetic monopole. Indeed,
as the Maxwell field generated by a Skyrmion is very small (for the
parameters fitting the actual mass of the nucleons) the external field is 
much larger than the Skyrmion's magnetic field. 

It would be interesting to find the differences between the 
electromagnetic quantities obtained from the ungauged model, as in \cite{ANW}, 
and our $U(1)$ gauged model. We are not able to compute the 
solutions of the $U(1)$ gauged model for the physical value of the parameter 
$\lambda_0$ as this is too large, but as we now know that the influence
of the gauge field is very small, we can compute the latter {\em
perturbatively} around the (ungauged) Hedgehog as an induced field. This
enables the evalution of the energy correction and the induced magnetic
moment. This perturbative analysis will be be carried out in the Section
{\bf 4}.

It can also be concluded that if the $U(1)$ gauged Skyrme model were
quantised as in \cite{ANW} (by quantising the zero
modes corresponding to the global gauge transformation) but taking into
account the electromagnetic field generated classically by the Skyrmion,
the result would not differ very markedly from what was obtained in
\cite{ANW}.

\subsection{$A_0 \neq 0$ : charged $U(1)$ Skyrmion}

We can now look for solutions with a non-zero electric charge by requiring
that the field $b$ in our ansatz \r{1.2} does not vanish. To do this we follow
the same procedure as Julia and Zee~\cite{JZ} and require that the electric 
field be
asymptotically of the form $A_0 = V_0 + q/(r^2+z^2)^{1/2}$ where $V_0$ and 
$q$ are two 
constants. In practice, one computes solutions for different values of $V_0$ 
and evaluates $q$ by computing the electric flux. We sought only those
solutions, for which the electric flux equals $4\pi$ times the charge of the 
electron.

It is important to realise that in this case, equations \r{eqgsk} are 
obtained after minimising the action and thus they do not minimise the
Hamiltonian \r{ham}.

In our units, the charge of the electron is $0.303$. 
In Figure 4 we show the energy as a function of $\lambda_0$, as well as 
$V_0$ as a function of $\lambda_0$, so that $q = 0.303$.

One can see that, for a fixed value of $\lambda_0$, the energy of the 
{\it charged gauged} Skyrmion is smaller than the energy of the {\it ungauged} 
Skyrmion when $\lambda_0 < 7$ but it is always larger than the energy of the 
{\it uncharged gauged} Skyrmion. If the energies of the electrically charged 
and uncharged gauged Skyrmions were interpreted as the the masses of the 
Proton and the Neutron $m_P$ and $m_N$, then on this purely classical level 
we would have to conclude that $(m_P -m_N) >0$ which is not correct. 
This is expected on the basis of its analogy with the dyon \cite{JZ}.
Clearly, 
to calculate this mass difference correctly one would have to perform the 
collective coordinate quantisation as in
Ref. \cite{ANW}, which we do not do here. 

\begin{figure*}[pt]
\unitlength1cm
\hfil\begin{picture}(6,8)
\epsfxsize=6cm
\epsffile{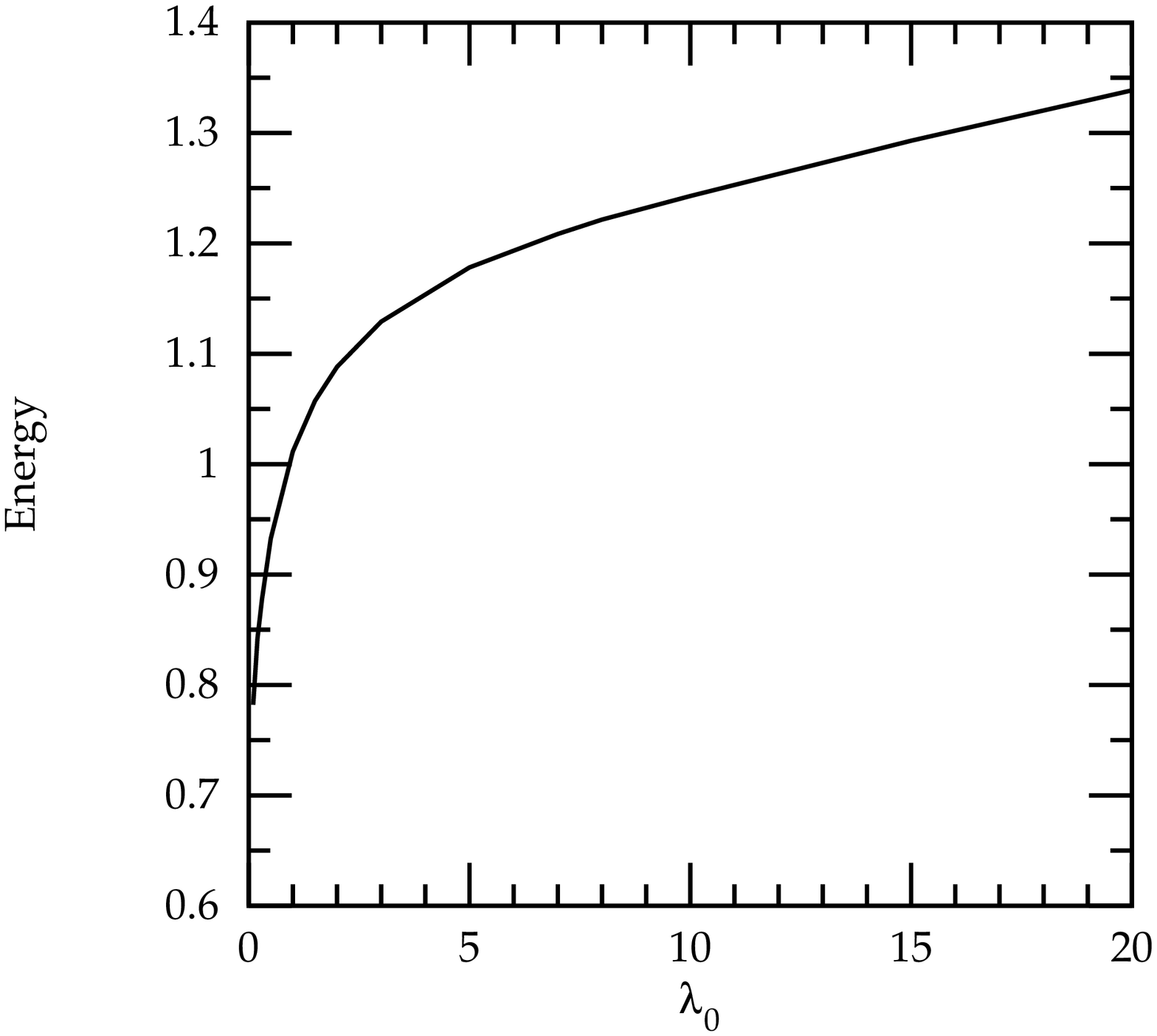}
\end{picture}
\hfil\begin{picture}(6,8)
\epsfxsize=6cm
\epsffile{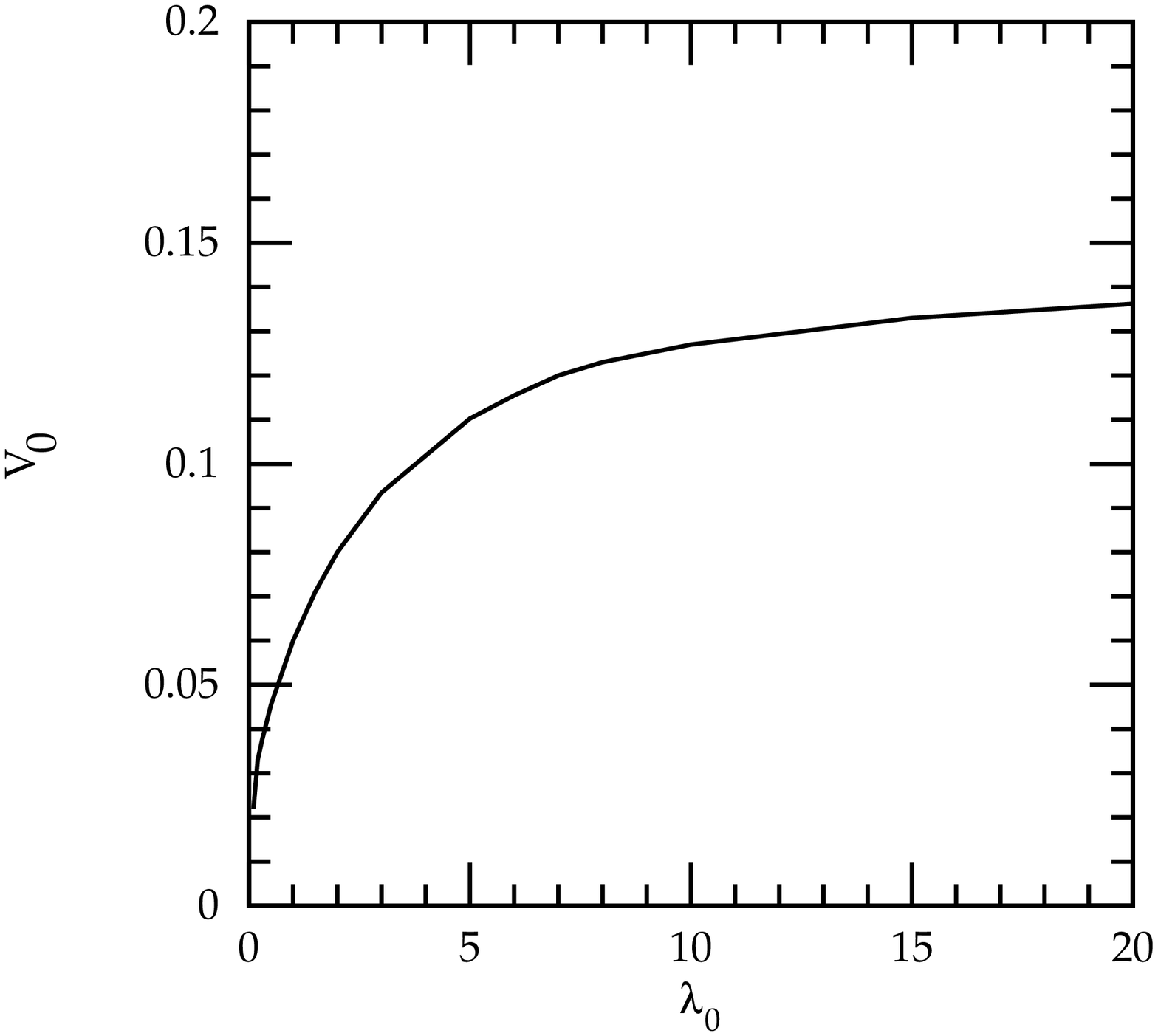}
\end{picture}
\caption{a) Energy of the charged Skyrmion 
b) $V_0$ as a function of $\lambda_0$}
\end{figure*}

The energy of the charged Skyrmion increases with $\lambda_0$. 
It is unfortunately very difficult to carry out the numerical computations
accurately when $\lambda_0$ is very large.  

At this stage, it is worth saying a few words about the numerical methods we 
have used. 
To compute the static solutions, we have employed a relaxation method using 
finite differences on a regular grid ($dr = dz$). This discretisation method
is similar to the one employed in the numerical computation of the
solutions of Skyrme  models in 2+1 and 3+1 dimensions \cite{BS,PZ,GH}.  
To compute the electrically charged solutions, we have imposed the boundary 
condition 
$b(\infty,\infty) = V_0$ for different $V_0$ and using a dichotomic method, 
we have determined the values that give a solution with the same electric 
flux as 
the proton. Most of the simulations where done on $200 \times 400$ or 
$300 \times 600$ grids. By computing the same solution for various lattice 
sizes, we have empirically obtained the following relation for the expression
of the energy of a solution : $E = E_0 + K dr^2$ where $E_0$ is the exact
solution, $dr=dz$ the lattice spacing and $K$ is a constant which depends on 
$\lambda_0$ but takes values between $1$ and $0.5$. We see thus that the 
energies we have obtained are accurate to within one or one half of a percent.
This inaccuracy in the value of the energy is comparable to that of
many other similar works on 2 dimensional systems \cite{PZ,GH}, and though it 
might look large, it does not affect any of the conclusions we have drawn.

\section{Perturbation around the Hedgehog}

We have seen from the work of {\bf 3.1} above that the energy of the $U(1)$
gauged Skyrmion for the physical values of the parameters, 
namely of the pion decay constant and the $U(1)$ coupling, does not differ
significantly from that of the ungauged Hedgehog. It is therefore
justified, for these values of the parameters, to treat the $U(1)$ field as
a perturbation to the Hedgehog in the same way as Klinhamer and
Manton~\cite{KM} treat the $U(1)$ field as a pertubation to the $SU(2)$
sphaleron. We will then be able to compute the magnetic moment of the 
Neutron, as well as the (small) deviation of its mass from that of the 
Hedgehog.

The equations for the fields $(\phi^a ,A_{\mu})$ are derived from the
Lagrangian (\ref{LagU}). The equation for $A_\mu$ will be of the form  
\begin{equation}
\label{lqe}
\lambda_0 \ \pa_\nu F_{\mu\nu} =j_\mu \ .\label{jcur}
\end{equation}
The method consists of setting the gauge field $A_\mu$ to zero in the current
$j_\mu$ in \r{jcur} (and in the equation for $\phi^a$) and calculating the
resulting induced electromagnetic field ${\rm A}_{\mu}$. The gauge field
computed this way can then be interpeted as the $U(1)$ field generated by the
(ungauged) Hedgehog Skyrme field. With this perturbative procedure, it is
possible to calculate the induced static magnetic potential ${\rm A}_i$
($i=1,2,3$), but not the static electric potential ${\rm A}_0$, which in this
scheme vanishes and can only be calculated non-perturbatively. The reason
simply is that restricting to the use of the static Hedgehog, the zeroth
component of the current $j_0$ at $A_{\mu}=0$ vanishes, resulting in turn in
vanishing induced potential ${\rm A}_0$ according to \r{jcur}.

As a consequence the electric field will be identicaly zero, which implies 
that we can derive the equation from the static Hamiltonian rather than
from the Lagrangian.
Notice also that we could try to compute perturbatively a solution for the 
electrically charges skyrmion by keeping in $j_0$ the terms proportional to 
$A_0$, instead of setting  $A_0$ to 0, and impose the condition that the 
electric field is assymptoticaly like that of the proton. This pertubation 
method would not make much sense though as one would expect the electric 
field to be quite large close to the skyrmion.

The relevant energy functional is \r{21}, and the resulting equation arising
from the variation of the gauge field $A_i$ is
\begin{eqnarray}
\label{p1}
&  \lambda_0 \ \pa_j F_{ij} =j_i \label{p1a} \\
j_i =&-\vep^{\alpha \beta} \left( {1\over 2}\lambda_1
\phi^\beta D_i \phi^\alpha +\lambda_2
\left[(\pa_j |\phi_{\gamma}|^2)
D_i \phi^\alpha D_j \phi^\beta+
2\phi^\beta D_{[i}\phi^\alpha D_{j]}\phi^A D_j \phi^A\right] \right)
\label{p1b}.
\end{eqnarray}
We are concerned here with the case where $A_i =0$ \r{p1b} and
the chiral field $\phi^a =(\phi^\alpha ,\phi^3 ,\phi^4)$ in \r{p1b}
describes the Hedgehog. {\it i.e.}
\be
\label{hedge}
\phi^\alpha =\sin F(R) \: \hat x^\alpha \: , \quad
\phi^3 =\sin F(R) \: \hat x^3 \: , \quad \phi^4 =\cos F(R) \: ,
\ee
where now $\hat x^a =x^a /R$, with $R=\sqrt{r^2 +z^2}$. By virtue of equation
\r{p1a} the current \r{p1b}, given by \r{hedge} and $A_i =0$, will now
induce a (small) $U(1)$ field ${\rm A}_i$, with curvature ${\rm F}_{ij}=\pa_i
{\rm A}_j -\pa_j {\rm A}_i$.

The shift in the energy of the Hedgehog due to the induced $U(1)$ field
${\rm A}_i$ is
\be
\label{p2}
\Delta E =\int d^3 x \: (\lambda_0 {\rm F}_{ij}{\rm F}_{ij}+4{\rm A}_i \:
{\rm j}_i)\: ,
\ee
in which ${\rm j}_i =j_i (0)$ is the current \r{p1b} for $A_i =0$.
(Note that all quantities evaluated at $A_i=0$ are denoted by Roman
script, {\it e.g.} ${\rm j}_i =j_i (0)$, as well as the induced connection and
curvature ${\rm A}_i$ and ${\rm F}_{ij}$.)
When equation \r{p1a} is satified for the induced $U(1)$ field,
\begin{eqnarray}
\Delta E &=&-\lambda_0 \int d^3 x \: {\rm F}_{ij}{\rm F}_{ij} \label{p3a} \\
&=& 2\int d^3 x \: {\rm A}_i \: {\rm j}_i \: ,\label{p3b}
\end{eqnarray}
which is, as expected, a strictly negative quantity.

The current
${\rm j}_i=({\rm j}_{\alpha},{\rm j}_3)$ in \r{p2} and \r{p3b} is given, for
the Hedghog field configuration \r{hedge}, by
\begin{eqnarray}
{\rm j}_\alpha &=&\frac{\sin^2 F}{R}\Bigg(
\frac{\lambda_1}{2}+2\lambda_2 \left(F'^2 +\frac{\sin^2 F}{R^2}\right)
\Bigg) \vep_{\alpha \beta}\hat x_\beta \label{p4a} \\
{\rm j}_3 &=&0\: .\label{p4b}
\end{eqnarray}

We now note that $\pa_i \ {\rm j}_i =0$, which means that equation \r{p1},
for $A_i=0$, takes the
following form in terms of the induced $U(1)$ connection
${\rm A}_i=({\rm A}_{\alpha},0)$
\be
\label{p11}
\lambda_0 \Delta \ {\rm A}_{\alpha} =-{\rm j}_{\alpha} \ .
\ee
The solution is well known and can be expressed, using the obvious notation
${\rm j}_{\alpha}({\bf x})={\rm j}(R)\ \vep_{\alpha \beta}\hat x_\beta$ in terms of
\r{p4a}, as
\be
\label{p12}
{\rm A}_{\alpha}({\bf x})=-\frac{1}{4\pi \lambda_0}\vep_{\alpha \beta}\int
\frac{1}{|{\bf x}-{\bf x'}|}\ {\rm j}(R')\ \hat x_{\beta}
\ d {\bf x'} \ ,
\ee
with~\cite{J}
\[
\frac{1}{|{\bf x}-{\bf x'}|}=\sum_{\ell =0}^{\infty}
\sum_{m=-\ell}^{\ell} \frac{4\pi}{2\ell +1}
\frac{R_<^{\ell}}{R_>^{\ell +1}}\ \bar Y^{(\ell)}_m (\theta' ,\phi') \ 
Y^{(\ell)}_m (\theta ,\phi) \ .
\]

After performing the angular integrations we have
\be
\label{p13}
{\rm A}_{\alpha}({\bf x})=-I(R)\ \vep_{\alpha \beta} \hat x_{\beta} \
\ ,
\ee
with $I_{(\ell)}(R)$ given by the integral
\be
\label{p13a}
I(R)=\frac{1}{3\lambda_0}\left(
\int_0^R \frac{R'}{R^2}
\: \: {\rm j}(R') \: R'^2 dR' +\int_R^{\infty}
\frac{R}{R'^2}
\: \: {\rm j}(R') \: R'^2 dR'  \right) \ .
\ee
Finally, in the $R\gg 1$ region of interest, the induced $U(1)$ potential is
\be
\label{p14}
{\rm A}_{\alpha}({\bf x})=-\frac{\hat I}{R^2}\ \vep_{\alpha \beta}\
\hat x_{\beta}\ ,\ {\rm with}\ ,\ 
\hat I=\frac{1}{3\lambda_0}\int_0^{\infty} s^3\:
{\rm j}(s)\: ds \ ,
\ee
to be evaluated numerically using the numerically constructed hedgehog
profile function $F(x)$ \r{hedge}.

Comparing \r{p14} with the usual Maxwell potential of a magnetic dipole
$\mbox{\boldmath $\mu$}$
\[
{\bf A}({\bf x})=\frac{ \mbox{\boldmath $\mu$}
\times {\bf x}}{4\pi R^3} \ ,
\]
we find that $\mbox{\boldmath $\mu$}=(0,0,\mu)$ is
\be
\label{p15}
\mu=4\pi \hat I\ .
\ee

We can evaluate the magnetic moment (\ref{p15}) and the energy correction 
(\ref{p3b}) induced  by the electromagnetic field by evaluating the integral 
(\ref{p13a}) and (\ref{p14}) numerically. If we take the experimental values 
$F_{\pi} = 186MeV$ and $a = 7.2$ we obtain 
\begin{eqnarray}
\mu &=& 0.01393 fm = 0.43 nm\\
\Delta E &=& -0.1 keV.
\end{eqnarray}
The experimental value for the magnetic moment of the proton and the neutron 
are respectivly 
$\mu_{p} = 0.0902 fm = 2.79 nm$ and $\mu_{n} = 0.0617 fm = -1.91 nm$.
If on the other hand we take the values of the parameters derived in 
\cite{ANW}, $F_{\pi} = 129MeV$ and $a = 5.45$, we have
\begin{eqnarray}
\mu &=& 0.0468 fm = 1.449 nm\\
\Delta E &=& -0.32 keV.
\end{eqnarray}

The magnetic moment of a particle is strictly speaking a quantum property 
and it should be computed by quantising the $SU(2)$ gauge degree of freedom 
as in \cite{ANW}. Nevertheless, we see that if the take the parameters
derived in \cite{ANW} the classical magnetic moment is of the correct 
order of magnitude. The sign is of course undetermined as the
classical magnetic moment is a vector. We can thus conclude that our model
offers a reasonable classical description of Nucleons and affords a method
for computing the electromagnetic field generated by the Skyrmion, classically.
It is quite surprising to see that a quantum property like the magnetic moment
can be reasonably predicted by a purely classical procedure.

\section{Summary and discussion}

We have shown that the $SU(2)$ Skyrme model gauged with $U(1)$ has two
types of finite energy static solutions, electrically uncharged and
charged respectively. Both of these solutions are axially symmetric
and carry no magnetic charge but support a magnetic field shaped like
a torus centred around the axis of symmetry, albeit resulting in zero
magnetic flux. The uncharged solutions, like the ungauged Skyrmion,
have a topological lower bound. The electrically charged solutions are the 
analogues of the Julia-Zee dyons~\cite{JZ} of the Georgi-Glashow model.

Concerning the stability of the electrically neutral solution, which is
expected to be stable by virtue of the lower bound on the energy, we have not
made any quantitative effort to test it. We expect however that the solitons
of this gauged Skyrme model are stable, or that at least they have stable branches
for all values of the parameters in the model. This expectation is based on our
knowledge of the corresponding situation when the Skyrme model is gauged instead
with $SO(3)$~\cite{BT,BKT}, in which case the equations arising from the
imposition of spherical symmetry were one dimensional, and hence technically
much more amenable to the numerical integration. In that case it was found
that in addition to stable branches of solutions, there were also some
unstable branches bifurcating from the former, the important matter being
that there were indeed stable branches of solutions, characterised
by the (ranges of the) parameters of the model. It would be very interesting to
carry out the analysis corresponding to that of \cite{BT,BKT}, for the
considerably more complex case of the axially symmetric equations at hand.
This however is technically beyond the scope of the present work.

The energies of the gauged uncharged Skyrmions are smaller than the energy of 
the usual ungauged Skyrmion. When the gauge coupling $1/\lambda_0^{1/2} $
goes to $0$, the uncharged gauged Skyrmion tends to the ungauged Skyrmion. 
We also note that the energy of the electrically charged Skyrmion is 
higher than the uncharged one, just as the mass of the dyon is higher than 
that of the monopole of the Georgi-Glashow model.

Perhaps the most interesting physical result of the present work is that
when parameters in the model are fitted to reproduce physical 
quantities, it turns out that
the effect of the Maxwell term in the Skyrme Lagrangian is very small.
This is because for the physical value of the constant $\lambda_0 = 1138$,
the energy of the gauged uncharged Skyrmion differs little from that of the 
ungauged Skyrmion, as seen from Figure 1b. The gauged Skyrmion field itself 
is thus nearly radially symmetric (though the gauge field is not). 

Having found that the influence of the electromagnetic field on the Skyrmion
is small, we were pointed in the direction of treating the magnetic potential
as an induced field perturbatively around the (ungauged) Hedgehog. We have been 
able thus, to compute the classical magnetic moment of the (uncharged) Skyrmion
of {\it unit} Baryon charge, namely of the Neutron.
The result is that the classical magnetic moment of the Skyrmion 
matches surprisingly well to the experimental values of the 
magnetic moments of the Nucleons. 

\bigskip

{\bf Acknowledgements} We are grateful to Y. Brihaye and B. Kleihaus for 
enlightening discussions , and especially to J. Kunz for having suggested to 
us the perturbative analysis in Section {\bf 4}. 
This work was supported under projects BCA 96/024 of
Forbairt/British Council, as well as IC/99/070 and IC/98/035 of
Enterprise--Ireland.

\end{document}